\documentclass[aip,amsmath,amssymb, reprint]{revtex4-2}

\usepackage{graphicx}
\usepackage{dcolumn}
\usepackage{bm}
\usepackage{upgreek}
\usepackage[utf8]{inputenc}
\usepackage[T1]{fontenc}
\usepackage{mathptmx}
\usepackage{etoolbox}
\usepackage{hyperref}
\usepackage{color}

\newcommand{\ch}[1]{{#1}}

\makeatletter
\def\@email#1#2{%
 \endgroup
 \patchcmd{\titleblock@produce}
  {\frontmatter@RRAPformat}
  {\frontmatter@RRAPformat{\produce@RRAP{*#1\href{mailto:#2}{#2}}}\frontmatter@RRAPformat}
  {}{}
}%
\makeatother
\begin{document}

\preprint{AIP/123-QED}

\title[Stand-alone vacuum cell for compact ultracold quantum technologies]{Stand-alone vacuum cell for compact ultracold quantum technologies}

\author{Oliver S.\ Burrow}
\affiliation{Department of Physics, SUPA, University of Strathclyde, Glasgow, G4 0NG, United Kingdom}
\author{Paul F.\ Osborn}
\affiliation{TMD Technologies Ltd, Swallowfield Way, Hayes UB3 1DQ, United Kingdom}
\author{Edward Boughton}
\affiliation{TMD Technologies Ltd, Swallowfield Way, Hayes UB3 1DQ, United Kingdom}
\author{Francesco Mirando}
\affiliation{Kelvin Nanotechnology Ltd., 70 Oakfield Ave, Glasgow, G12 8LS, United Kingdom}
\author{David P.\ Burt}
\affiliation{Kelvin Nanotechnology Ltd., 70 Oakfield Ave, Glasgow, G12 8LS, United Kingdom}
\author{Paul F.\ Griffin}
\affiliation{Department of Physics, SUPA, University of Strathclyde, Glasgow, G4 0NG, United Kingdom}
\author{Aidan S.\ Arnold} \email{aidan.arnold@strath.ac.uk}
\affiliation{Department of Physics, SUPA, University of Strathclyde, Glasgow, G4 0NG, United Kingdom}
\author{Erling Riis}
\affiliation{Department of Physics, SUPA, University of Strathclyde, Glasgow, G4 0NG, United Kingdom}

\date{\today}

\begin{abstract}
Compact vacuum systems are key enabling components for cold atom technologies, facilitating extremely accurate sensing applications. There has been important progress towards a truly portable compact vacuum system, however size, weight and power consumption can be prohibitively large, optical access may be limited, and active pumping is often required. 
Here, we present a centilitre-scale ceramic vacuum chamber with He-impermeable viewports and an integrated diffractive optic,  enabling robust laser cooling with light from a single polarization-maintaining fibre. A cold atom demonstrator based on the vacuum cell delivers $10^7$ laser-cooled $^{87}$Rb atoms per second, using minimal electrical power. With continuous Rb gas emission active pumping yields a $10^{-7}\,$mbar equilibrium pressure, and passive pumping stabilises to $3\times 10^{-6}\,$mbar, with a $17\,$day time constant. A vacuum cell, with no Rb dispensing and only passive pumping, has currently kept a similar pressure for more than \ch{500 days}. The passive-pumping vacuum lifetime is several years, estimated from short-term He throughput, with many foreseeable improvements. This technology enables wide-ranging mobilization of ultracold quantum metrology.
\end{abstract}

\maketitle


The ability to laser cool atoms yields orders of magnitude longer interrogation times than with room temperature atoms for equivalent volume devices, enabling measurements with unprecedented accuracy \cite{Dutta2016,mcgrew2018atomic,PhysRevLett.120.183604}. Whilst the most precise of these instruments are room-sized apparatus, a generation of compact quantum technologies are being developed \cite{bidel2018absolute,liu_-orbit_2018,becker_space-borne_2018,Grotti2018,aveline_observation_2020,Takamoto2020} 
to take ultracold atomic accuracy out of the lab and into real-world applications. The key to achieving this goal is to reduce the size, weight and power (SWaP) of the device's individual components, whilst increasing simplicity and resilience.

Quantum cold atom sensors use magneto-optical traps (MOTs)  \cite{Raab1987,Monroe1990,Barry2014} 
comprising laser-cooling sub-systems of: lasers, magnetic coils, optics, and \ch{sufficient} vacuum. 
Development of compact laser systems is a subject of on-going research, using diode laser and telecommunications industry technology for robust miniaturisation  \cite{lienhart2007compact,San2012,Lvque2014,Pahl2019,DiGaetano2020}. 
Magnetic coils optimised for low power consumption can be designed and fabricated \footnote{Private communication: Nathan Welch and Mark Fromhold, University of Nottingham.}, and the optics for laser cooling can be simplified to a single beam illuminating a pyramidal or planar optic \cite{Lee1996,Vangeleyn2009,Vangeleyn2010}. However, while key progress has been made in developing miniaturised vacuum systems for `hot' ions \cite{Jau2012,Schwindt2016} and laser-cooled atoms \cite{Scherer2012,himsworthreview,Basu2016,McGilligan2020APL,Little2021}, a chamber should ideally be devoid of any challenging bulky components or appendages, with integrated pump and atom source, enabling \ch{suitable} vacuum in an apparatus with a truly compact form factor.

Custom components have been required to improve on the vacuum system SWaP, leading to vacuum seal challenges, particularly when including the necessary optical access.  Ideally the system would be passively pumped \cite{Scherer2012,Boudot2020}, to eradicate vacuum power requirements.
The resulting finite vacuum lifetime could be maximised by careful choice of materials to minimise outgassing and the permeation of non-pumpable noble gases \cite{himsworthreview,Dellis16}. Passive pumping also means that the undesirable volume and magnetic field of an ion pump can be removed, ameliorating Zeeman systematic shifts on precision atomic measurements.

\begin{figure*}[!t]
\centering
\includegraphics[width=\textwidth]{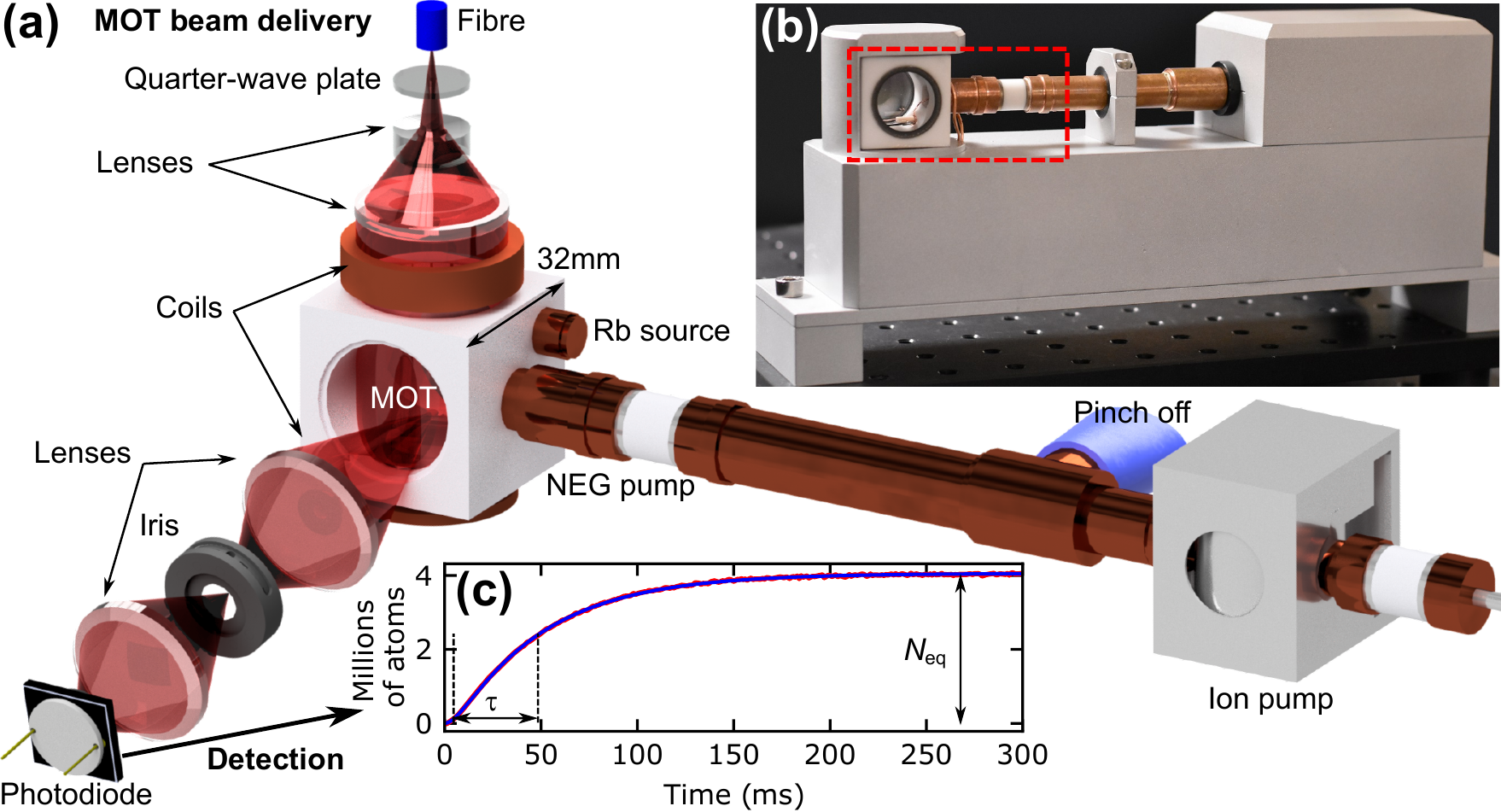}
\caption{The portable vacuum chamber assembly (a), including beam delivery, detection and magnetic systems. A photograph of the complete demonstrator chamber, with all delivery optics and electronics is depicted in (b), with the essential passively-pumped vacuum footprint for future devices highlighted in red. An experimental Rb MOT loading curve is shown in (c).}
\label{fig:cellphoto}
\hspace{5.00mm}
\end{figure*}

Here, we present a cubic high vacuum chamber, with $32\,$mm sides, and an integrated diffraction grating for magneto-optical trapping of atoms with a single input laser beam: a grating MOT (gMOT \cite{Nshii2013,mcgilligan2017grating,Elvin2019}) vacuum cell (Fig.~\ref{fig:cellphoto} a). With active pumping the cell pressure is $10^{-7}\,$mbar, measured via the MOT loading curves, and maintained indefinitely -- ideally suited for fast operation as an ultracold atom source for metrology experiments.
With purely passive pumping, the pressure rises to an asymptote of $3\times 10^{-6}\,$mbar,  
even with a relvatively impure atom dispenser continuously running.
A second vacuum cell has maintained a similar pressure with no active pumping or Rb dispensing for \ch{500 days}.

Weighing in at $252\,$g the vacuum assembly  (Fig.~\ref{fig:cellphoto} (a)) has been used in a portable demonstrator (Fig.~\ref{fig:cellphoto} (b)), laser cooling $10^6$ atoms at $10\,$Hz repetition rate, with  $20\,$mW of gMOT optical power delivered by fibre. Electrical power of $\approx 10\,$W is sufficient for Rb dispensers, ion pump and anti-Helmholtz coils. A  printed circuit board, with a USB $5\,$V input, can run the whole demonstrator with the required voltages for the  dispensers, coils  and  ion  pump, and a USB battery pack then provides several hours of use. The vacuum cell's $25\times4.0\times3.2\,$cm$^3$ volume could be substantially reduced if the redundant ion pump (the only magnetic component) is removed.
This vacuum cell with no active pumping,  $\approx10^{-6}\,$mbar pressures and projected helium-limited lifetime of several years will be enhanced in future via a new generation of `cleaner' atomic reservoirs \cite{Kohn2020,McGilligan2020,Li2019}.


To realise a portable magneto-optical trap, we developed a vacuum enclosure to fulfill the  necessary requirements (Fig.~\ref{fig:cellphoto}): good optical access, high vacuum pressures  $\approx 10^{-6}\,$mbar, and  a controllable source of Rb vapour. 
Our system is based around the gMOT architecture, where the grating properties\cite{McGilligan2016,Cotter2016} and gMOT phase space density\cite{Nshii2013,McGilligan2015,mcgilligan2017grating,Elvin2019} were fully characterised\footnote{The timescales for vacuum collisions are many orders of magnitude longer than the timescale of the laser cooling dynamics.} in earlier work. A single input beam and a single $2\times2\,$cm$^2$ in-vacuo micro-fabricated diffractive optic (patterned with e.g.\ 1D  gratings) are used to cool $\approx 10^7$ atoms \cite{Nshii2013}, with $3\,\upmu$K temperature demonstrated  \cite{mcgilligan2017grating} for atom populations similar to ours here.

To achieve the required vacuum, components must be chosen which have low outgassing rates and helium permeability. Materials and bonding methods used must also withstand 
a high temperature vacuum bake in the cell construction and subsequent outgassing.
The gMOT vacuum cell comprises an OFE-copper pump stem assembly bonded to a sintered alumina cubic chamber with a high temperature metallic braze. The pump stem assembly incorporates a Ti cathode sputter-ion pump  
and a non-evaporable getter pump (NEG) \footnote{The ion pump has nominal speed $0.2\,$L/s and the NEG is the electrically-activated SAES Getters St-172. We suspect  $5$-$10$ times the NEG material would be useful.}.  
The stem assembly was attached to a vacuum system for high temperature vacuum bake via a standard DN40CF flange, and then subsequently removed by a pinch-off technique cold-welding the OFE-copper tube closed under application of mechanical pressure.

The cubic chamber has $32\,$mm side dimensions, and houses the optical grating. A glass-ceramic optical viewport is oriented facing the grating to provide access for the trapping laser, with two further windows on adjacent sides for probe laser access and observation (Fig.~\ref{fig:cellphoto} (a), (b)). The windows are attached using an established glass-bonding technique.
An OFE-copper port adjacent to the pump stem allows a SAES Getters rubidium dispenser to be positioned in the alumina cube, with electrical contact to the dispenser made between the port and the pump stem. Vacuum bake is performed $\geq 300 ^\circ$C for 100 hours, during which the Rb dispenser and NEG are activated. 


To facilitate characterisation a testing rig was constructed which receives the vacuum cell, mechanically centring and aligning it to both the trapping laser beam and magnetic field. The rig has a pair of anti-Helmholtz type coils providing a quadrupole magnetic field with $18\,$G/cm axial gradient, and three pairs of Helmholtz coils to cancel background magnetic fields. The rig's delivery beam optic assembly receives a polarisation-maintaining optical fibre, expands and collimates the beam to a $1/e^2$ radius of $15\,$mm and circularly polarises the light. The fibre delivers 
$780\,$nm light to the beam delivery system, red-detuned $12\,$MHz from the $F = 2 \rightarrow F'=3$ $^{87}$Rb D2 transition, with a repumper sideband of  $\approx1\%$ power  \cite{Vangeleyn2010} provided by a $6.58\,$GHz electro-optical modulator.  

Compared to the atomic fluorescence, the light scattered in the cell can be high, due to diffracted light from the grating hitting the cell walls. A clean signal from the atoms is detected by a photodiode in a spatial filtering lens system \cite{McGilligan2020APL}, minimising light not originating from the atom trapping point (Fig.~\ref{fig:cellphoto} (a)). 
Loading curve data (Fig.~\ref{fig:cellphoto}(c)) was taken at $1\,$ms resolution with a $2\,$s MOT cycle time. The MOT was turned off and on using the quadrupole magnetic field, with coil switching time $<2\,$ms, and an `on' time of $0.5\,$s.

To characterise the cell's vacuum performance, the pressure must be measured.  Ion gauges are typically used to measure pressure in the desired range, however their bulk renders them impractical for a compact system.
The ion pump current can be used to measure pressure, however the current-pressure relationship varies from pump to pump and leakage currents 
 provide a species-specific systematic error. 
The ion pump will be removed from future system designs, which would also remove this capability.  Furthermore, ion gauges and pumps do not measure the pressure at the MOT position, i.e.\ the vacuum conductance between locations must be accurately known.  

The pressure at the MOT location can instead be reliably measured using the atom loading dynamics \cite{PhysRevA.85.033420,Moore2015,Scherschligt2018}.  It has long been established that the pressure affects the lifetime of an atom in a cold atom trap, and this relationship can be used in reverse to measure the pressure \cite{Monroe1990}.  The equation for the derivative of MOT atom number $N$  at time $t$ is:
\begin{equation}
\ch{\dot{N}(t)} = \alpha \, P_\textrm{Rb} - (\beta\, P_\textrm{Rb} + \gamma\,  P_\textrm{bk})\, N(t),
\end{equation}
where MOT loading is proportional to the rubidium vapour pressure $P_\textrm{Rb}$, and the coefficient $\alpha$ is specific to the experimental apparatus and parameters.
Cold atoms exit the MOT at two rates: due to collisions with the Rb vapor or other non-rubidium background gases at pressures $P_\textrm{Rb}$ and $P_\textrm{bk}$, with corresponding loss coefficients\cite{PhysRevA.85.033420} $\{\beta,\gamma\} = \{3.3,3.7\} \times 10^{7}\,\textrm{mbar}^{-1}\,\textrm{s}^{-1}$.  The non-rubidium gases are assumed to be dominated by H$_2$ \ch{(shown to be the case by residual gas analyser measurements at the end of the bake)}, with the cross-sections of other species varying by at most a factor of two\ch{\cite{PhysRevA.85.033420}}. Two- and three-body rubidium collisions are also assumed to be negligible.

A brief delay, of a few ms, was observed between initiating laser cooling and measuring the fluorescence signal from trapped atoms.  As the imaging system here is only sensitive to atoms at the MOT position, this is due to the time taken for the atoms to cool and trap from the beam overlap region and then congregate at the trap centre. A  simulation of forces in our atom trapping volume and the imaging region, yielded comparable values (a few ms) to the observed trapping time. This effect was empirically modelled by replacing the constant $\alpha$, with $\alpha(t) = \alpha(1 - e^{-t/\delta}),$
where $\delta$ is the characteristic time for atoms entering the trap volume to arrive at the imaging location.  

Using this delay time and the following relations for the measurable quantities, the MOT's equilibrium atom number and lifetime are:
\begin{equation}
N_\textrm{eq} = \alpha \, P_\textrm{Rb} \, \tau,~~ \tau^{-1} = \beta \, P_\textrm{Rb} + \gamma \, P_\textrm{bk},
\label{eq:Neq}
\end{equation} 
respectively, with solution
\begin{equation}
N(t) = N_\textrm{eq} \bigg(1 + \frac{\delta \, e^{-t/\delta} - \tau \, e^{-t/\tau} }{\tau - \delta}\bigg).
\label{eq:N_Erling}
\end{equation}
This recovers the result from \cite{Moore2015} when $\delta\rightarrow 0$. 
This slight modification provides a good fit to the measured \ch{background-subtracted} fluorescence (Fig.~\ref{fig:cellphoto}(c)), preventing overestimation of the equilibrium atom number \footnote{Similar results are also obtained by fitting \ch{the background-subtracted fluorescence data with $(N_\textrm{eq}-N_1) (1-e^{-t/\tau})+N_1$, to determine $N_\textrm{eq}$.  Data at early times is omitted to find $\tau$, and $N_1$ is solely a fit parameter.}}.

\begin{figure}[!b]
\includegraphics[width=\columnwidth]{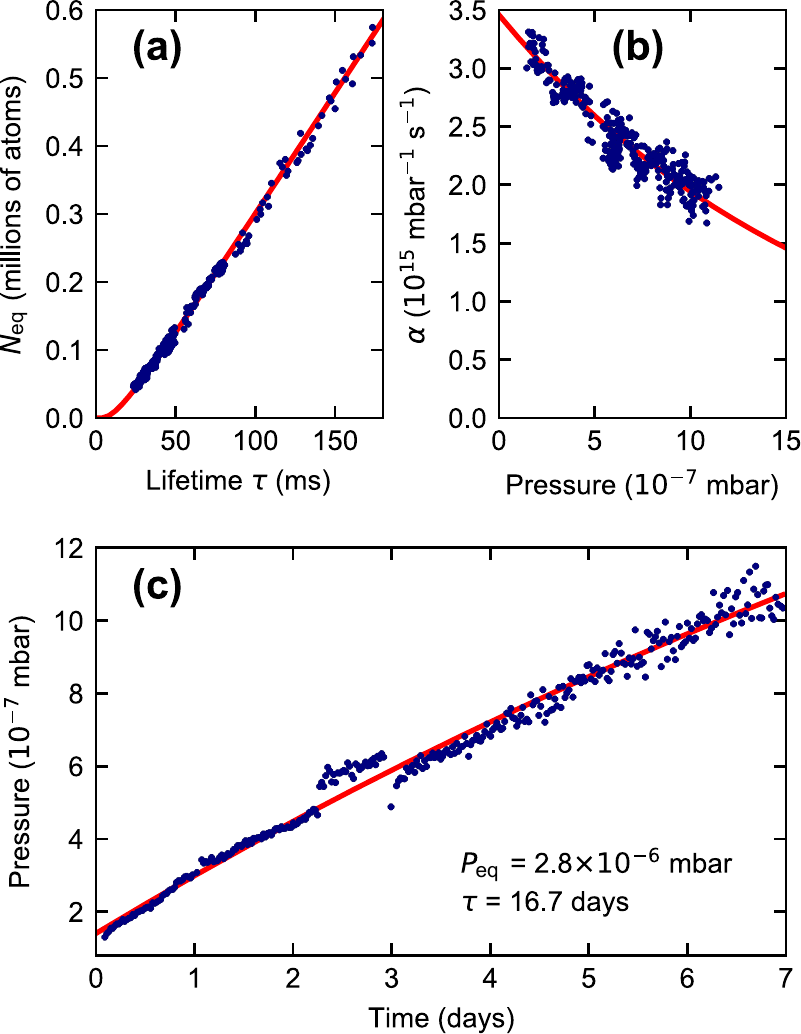}
\caption{The evolution of MOT atom number $N_\textrm{eq}$ vs.\ lifetime $\tau$  in one week of continuous loading curve measurements, with only passive pumping and constant Rb gas dispensing (a). The ratio $N_\textrm{eq}/\tau$ is proportional to the trap loading rate $\alpha P_\textrm{Rb}$ (b). The pressure rises when the ion pump is turned off, and the rate atoms are loaded into the trap decreases when the atom loss due to collisions becomes comparable with the fitted MOT capture time \ch{$\tau_\textrm{C}=16\,$ms} used via Eq.~\ref{Eq:Fig2ab} for the theory curves in (a) and (b). From these results and Eq.~\ref{eq:Neq} one can infer the background gas pressure against time (c) during the week of passive pumping and Rb dispensing.}
\label{fig:AlphaVsP}
\end{figure}

\ch{Over the course of a week, with only passive pumping and a constant Rb dispenser current \footnote{We used a lower current for the dispensers than the data shown in Fig.~\ref{fig:cellphoto}(a) to keep $P_\textrm{Rb}$ low and have minimal error on $P_\textrm{bk}$ in Eq.~\ref{eq:Neq}.}, we studied the MOT fill parameter behaviour  $N_\textrm{eq}$ vs $\tau$, Fig.~2(a). These fill parameters are accurately determined by fitting MOT loading curves with Eq.~\ref{eq:N_Erling}, however there are three unknowns in Eq.~\ref{eq:Neq}: $P_\textrm{Rb}$, $P_\textrm{bk}$ and the experiment-specific $\alpha$.  Whilst the procedure laid out in Ref.~\cite{Moore2015} can be followed to measure $\alpha$, and hence $P_\textrm{Rb}$ and $P_\textrm{bk}$, we found that the quantity $\alpha P_\textrm{Rb}=N_\textrm{eq}/\tau$ actually varied with $\tau$. The behaviour of $N_\textrm{eq}/\tau$ vs.\ $1/\tau$ was very similar to Fig.~\ref{fig:AlphaVsP}(b) and inconsistent with our expectation that $P_\textrm{Rb}$ remained constant. Using low-intensity $780\,$nm absorption spectroscopy of the MOT vapor cell \cite{elecsus2}, assuming a uniform Rb distribution, we independently confirmed a constant $P_\textrm{Rb}=2\times 10^{-9}\,$mbar. 
Using this constant $P_\textrm{Rb}$ in conjunction with Eq.~\ref{eq:Neq} we directly determined  $\alpha=N_\textrm{eq} (\tau P_\textrm{Rb})^{-1}$ and its variation with $P_\textrm{bk}=(\gamma \tau)^{-1}-\beta \gamma^{-1} P_\textrm{Rb}$,   illustrated in Fig.~\ref{fig:AlphaVsP}b.  

We have therefore found} evidence justifying an improved MOT loading model with pressure-dependent $\alpha$. Moreover, \ch{the physical reason for this is} the increasing importance of background collisions which occur on timescales comparable to the MOT characteristic capture time $\tau_\textrm{C}$ 
\footnote{This capture time is in reasonable agreement with a simple Doppler cooling model \cite{McGilligan2015} that has been enhanced using stochastic processes.}\ch{. One} can obtain an excellent fit to the data \ch{in Fig.~\ref{fig:AlphaVsP}(b)} using a Poisson distribution and consider only atoms surviving the cooling process relative to the pressure-dependent collisional loss rate \footnote{Alternatively, constant $\alpha P_\textrm{Rb}$ can be more simply reconciled if one allows a $\tau$ offset for the minimum lifetime $\tau_\textrm{C}=16\,$ms at which MOTs are obtained. A linear fit in Fig.~\ref{fig:AlphaVsP}(a) gives $\alpha P_\textrm{Rb}=N_\textrm{eq}/(\tau-\tau_\textrm{C})$.}. In particular Fig.~\ref{fig:AlphaVsP} (a) and (b) are fit with the curves:
\begin{equation}
N_\textrm{eq}=\ch{\alpha_0}\,P_\textrm{Rb}\,\tau\,e^{-\tau_C/\tau},  \;\;\;\frac{N}{\,P_\textrm{Rb}\tau}=\alpha'\,e^{-\tau_C\,\gamma \,P_\textrm{bk}},
\label{Eq:Fig2ab}
\end{equation}
 respectively, where $\alpha'=\ch{\alpha_0}\,e^{-\tau_C\, \beta\, P_\textrm{Rb}}$ \ch{tends to $\alpha_0$} at low Rb pressures \footnote{We note that the pressure decay coefficient $P_0=(\tau_\textrm{C}\,\gamma)^{-1}$ links $\tau_\textrm{C}$ and $\gamma$ -- enabling a clear determination of one parameter provided independent measurement of the other}.

Utilising the equilibrium MOT atom number $N_\textrm{eq}$ and time constant $\tau$ to determine pressure via Eq.~\ref{eq:Neq}, we have explored the behaviour of the vacuum chamber under a variety of conditions. Following the characterisation of $N_\textrm{eq}$-$\tau$ evolution of the cell with the ion pump off and dispensers on in (Fig.~\ref{fig:AlphaVsP} (a), (b)), we could determine the background pressure $P_\textrm{Rb}$ rise over the course of the week. Fitting this curve with a function \begin{equation}
P(t) = P_0+(P_\textrm{eq}-P_0) (1-e^{-(t-t_0)/T_\textrm{V}}),\nonumber
\end{equation}
yields the parameters for asymptotic equilibrium pressure $P_\textrm{eq}=3\times 10^{-6}\,$mbar and rise time constant $T_\textrm{V}=17\,$days.  

This observation of passive pumping reaching equilibrium is supported by a pressure measurement in a similar cell of \ch{$2.5 \times 10^{-6}\,$mbar after 500 days} with no active pumping during which no Rb was dispensed.  \ch{Furthermore, a cell has been pinched off entirely from its ion pump and is following a similar pressure trajectory to one with only passive pumping.} 
We largely attribute the pressure rise in  Fig.~\ref{fig:AlphaVsP} (c) to non-Rb emission from the dispensers, as the pressure ceases to rise dramatically when the dispensers are off for several hours, and the pressure load from the dispensers due to Rb is small -- $P_\textrm{Rb}=2\times 10^{-9}\,$mbar at equilibrium with passive-only or active pumping.


Noble gases are of particular concern when trying to eliminate active pumping from a vacuum system.  Passive pump mechanisms do not pump noble gases, and as such their throughput into the vacuum cell must be minimised.  Helium is of particular concern as the smallest mono-atomic gas readily leaks through any channels, and could permeate directly through the walls and windows of a poorly designed vacuum cell. Amongst its many exciting properties, graphene coating also offers prospects to prevent He permeability \cite{Sun2020}. The partial pressure of helium in Earth's atmosphere is $5\times 10^{-3}\,$mbar, meaning there is a strong pressure differential between this and the desired $1\times 10^{-6}\,$mbar overall pressure in the vacuum cell. 

We studied accelerated He permeation, by first placing the cell in equilibrium with Rb dispenser current and ion pump on, after which the dispensers and ion pump were turned off and the cell was disconnected from the testing rig ($t=0\,$minutes). The cell was then placed in a pressure vessel with $1\,$bar of He for 5 minutes, before being fully returned to the testing rig at $t=18\,$minutes. With only the dispenser current turned back on, and no ion pumping, the pressure change is $+2.3\times 10^{-7}$ mbar, whereas the maximum expected pressure rise established in Fig.~\ref{fig:AlphaVsP} would be negligible (and over-estimated as the dispensers are initially off). Attributing this pressure rise solely to the 5 minutes at $1\,$bar of helium, the time scale for a $10^{-6}\,$mbar pressure rise with atmospheric He ($5\times10^{-3}\,$mbar i.e. $5\,$ppm in pressure) would therefore already be 8 years. 


We have created a centilitre volume vacuum cell, with integrated grating-MOT optics, which can be used as a robust cold atom source. With active ion pumping the cell is expected to function for several years at $10^{-7}\,$mbar, based on the specified total pumping capability. 
With passive-only pumping the cell's pressure increases to a higher equilibrium pressure of $3\times 10^{-6}\,$mbar, with an exponential time constant of $T_\textrm{V}=17\,$days, 
but still has an expected lifetime of several years.  

The passive-pumping pressure rise is largely due to the Rb dispensers, which at above 500$^\circ$C are likely to be outgassing other \ch{non-rubidium} species significantly. Moreover, the dispenser is currently the main contributor to the system's electrical power. 
A different rubidium source would ameliorate the situation, e.g.\ lower temperature methods include capturing Rb in graphite, for either thermal \cite{Kohn2020} or electronic release \cite{McGilligan2020}. Microfabrication-based beam options are also now available \cite{Li2019}. Rubidium metal could in principle be used, as it has  $2\times10^{-7}\,$mbar room temperature vapor pressure, however it would have to be distilled to a cold point in the chamber, or released from a small ampoule, after the cell fabrication and bake process. 

Accelerated He permeation tests already indicate the passively-pumped cell can withstand eight years of atmospheric helium leakage, and finding the main channel for He permeation in future tests could extend this lifetime significantly. 
Whilst the cell does contain a non-evaporable getter pump, optimising the quantity and activation of this remains a fertile avenue of investigation, with potential for a large improvement.

This cell could be used immediately as the atom source for practicable quantum metrology.  Many cold atomic sensors operate with a measurement time in the 1-20$\,$ms range  \cite{Wu2017,Pelle2018,Rakholia2014,bidel2018absolute,Chen2019,Elgin2019}, 
which is already shorter than (and therefore largely unaffected by) our atomic collisional loss times. A concrete example of a device achievable using the cell presented here would be a compact cold-atom CPT clock \cite{Elvin2019} with repetition rate and Ramsey linewidth of $25\,$Hz. With $10^6$ atoms loaded in $20\,$ms  (Fig.~\ref{fig:cellphoto} (c)), followed by $20\,$ms interrogation in free-fall,  short term fractional-frequency stabilities of $10^{-12}\,\textrm{s}^{-1/2}$ would be possible  \cite{Santarelli1999,Esnault2011,Newman2019}.

Novel cell geometries could be fabricated with our techniques: smaller volumes; extended cell lengths for longer free-fall times; hardware/cavity incorporation for long interrogation times \cite{Abend2016,Xu2019}; and differentially pumped double-chambered vessels for  compact Bose-Einstein condensates  \cite{Straatsma2015}. 
Moreover, the cell could be adapted as a key component for demanding marine or space environments \cite{bidel2018absolute,liu_-orbit_2018,becker_space-borne_2018,Grotti2018,aveline_observation_2020},  used in quantum memories \cite{Tranter2018,Cao2020} and the diffractive element could facilitate a variety of single-input-beam optical lattice geometries \cite{Nshii2013} for optical lattice clocks \cite{Takamoto2020} or quantum simulators \cite{Bernien2017,Zhang2017}.

Our vacuum device will aid the mobilization of quantum technologies, across a diverse range of applications that require the accuracy that only cold atoms, ions and molecules provide. The data that support the findings of this study are openly available via the Strathclyde Pureportal  \footnote{Dataset DOI TBA}.

We are grateful for support through InnovateUK grant \href{https://gtr.ukri.org/projects?ref=103880}{103880: `gMOT - Magneto optical trap system for cold atom technologies'} (\href{https://gow.epsrc.ukri.org/NGBOViewGrant.aspx?GrantRef=EP/R020086/1}{EP/R020086/1}), as well as Quantum Technology Hub EPSRC grants \href{https://gow.epsrc.ukri.org/NGBOViewGrant.aspx?GrantRef=EP/T001046/1}{EP/T001046/1} and \href{https://gow.epsrc.ukri.org/NGBOViewGrant.aspx?GrantRef=EP/M013294/1}{EP/M013294/1}. We thank Alan Bregazzi and James McGilligan for guidance in the design of the imaging system and for useful discussions.

\nocite{*}
\bibliography{Bibliography}

\end{document}